# Impersonating a Superconductor: High-Pressure BaCoO$_3$, an Insulating Ferromagnet

Haozhe Wang[1‡], Xianghan Xu[2‡], Danrui Ni[2‡*], David Walker[3], Jie Li[4], Robert J. Cava[2], Weiwei Xie[1*]

1. Department of Chemistry, Michigan State University, East Lansing, Michigan, 48824
2. Department of Chemistry, Princeton University, Princeton, New Jersey, 08540
3. Lamont Doherty Earth Observatory, Columbia University, Palisades, New York, 10964
4. Department of Earth and Environmental Sciences, University of Michigan, Ann Arbor, Michigan, 48109

**ABSTRACT:** We report the high-pressure synthesis (6 GPa, 1200 °C) and ambient pressure characterization of hexagonal HP-BaCoO$_3$. The material (with the 2H crystal structure) has a short intrachain Co–Co distance of about 2.07 Å. Our magnetization investigation revealed robust diamagnetic behavior below approximately 130 K when exposed to weak applied magnetic fields (10 Oe) and a distinct "half-levitation" phenomenon below that temperature, such as is often observed for superconductors. Its field-dependent magnetization profile, however, unveils the characteristics of ferromagnetism, marked by a substantial magnetic retentivity of 0.22(1) $\mu_B$/Co at a temperature of 2 K. Electrical resistivity measurements indicate that HP-BaCoO$_3$ is a ferromagnetic insulator, not a superconductor.

Ferromagnetic materials can exhibit diamagnetic characteristics, such as partial ("half") levitation, when subjected to weak magnetic fields.[1] In a recent development, researchers from Korea reported the discovery of room-temperature superconductivity under atmospheric pressure in a copper-substituted lead phosphate apatite with the chemical composition Pb$_{10-x}$Cu$_x$(PO$_4$)$_6$O, also denoted as "LK-99".[2] This material was scrutinized at Peking University, revealing that LK-99 instead demonstrates the attributes of soft ferromagnetism, featuring substantial magnetic anisotropy that gives rise to the phenomenon of half-levitation.[3] However, it is worth noting that occurrences of such "half-levitation" phenomena are not uncommon in the realm of ferromagnetic materials.

Chemical design rules for finding magnetic materials are empirically based on the magnetic orbital interactions and the symmetry of the material. By tuning the inter-atomic distances and angles, the magnetic ordering can be manipulated. In addition, cobalt-based oxides have received extensive attention due to the diverse spin states that Co can display, which has been attributed to the comparable magnitudes of the crystal electric field (CEF) splitting of the Co d states and the Hund's rule exchange energy.[4,5,6] Consequently, the energy gap between the $t_{2g}$ and $e_g$ states is relatively small, enabling thermal excitation of $t_{2g}$ electrons to the $e_g$ states, leading to higher spin states and large magnetic moments.[7,8,9] Our 2H-BaCoO$_3$, for example, is a weak magnetic insulator with low spin $S = 1/2$ Co$^{4+}$.[10,11] In contrast, SrCoO$_3$ is characterized as a ferromagnetic (FM) metal with a Curie temperature ($T_C$) of ~280 K.[12,13]

2H-BaCoO$_3$ consists of parallel-to-$c$ chains of face-sharing CoO$_6$ octahedra that form a 2D triangular lattice in the $ab$-plane, in which the in-chain face-sharing CoO$_6$ octahedra connect neighboring Co$^{4+}$ with a Co$^{4+}$–O$^{2-}$–Co$^{4+}$ angle of about 78°.[10] Different from 2H-BaCoO$_3$, in contrast, 180° Co$^{4+}$–O$^{2-}$–Co$^{4+}$ exchange exists in three equivalent directions in the cubic perovskite SrCoO$_3$.[14,15] In addition to tuning the Co$^{4+}$–O$^{2-}$–Co$^{4+}$ angle, changing the Co–Co distance may also affect the magnetic properties.

Here we report the high-pressure synthesis, structural characterization, and a survey of the magnetic and electrical transport properties of the new hexagonal Co-based-oxide BaCoO$_3$. Magnetic susceptibility measurements show diamagnetic behavior at low magnetic fields but ferromagnetism at high magnetic fields. The material is weakly levitated below its magnetic ordering transition temperature. Electrical resistivity measurements demonstrate insulating properties at both room and low temperature.

The high-pressure hexagonal Co-based-oxide BaCoO$_3$ (HP-BaCoO$_3$) was synthesized by heating and pressurizing 2H-BaCoO$_3$ at 6 GPa. The HP-BaCoO$_3$ phase was confirmed to be in a very major proportion in the product, with some BaCoO$_{2.6}$ also present, as determined from thermogravimetry analysis (TGA) (**Figure S1**), and powder X-ray diffraction (PXRD). The PXRD patterns of BaCoO$_3$, analyzed by TOPAS before pressurization and HP-BaCoO$_3$ after pressurization, are presented in **Figure 1**. The differences in the PXRD patterns clearly show the peak shifting due to the difference in the lattice parameters before and after high pressure and high temperature treatment. Similar to 2H-BaCoO$_3$, HP-BaCoO$_3$ crystallizes in the hexagonal structure ($a$ = 5.648(1) Å, $c$ = 4.138(2) Å) with the space group $P6_3/mmc$ (#194), where all the nearest-neighbor CoO$_6$ octahedra are face-shared, with the Co–Co distance 2.069(1) Å, much shorter than the Co–Co distance of 2.3815(2) Å seen in ambient pressure BaCoO$_3$.[10] Thus, the high-pressure synthesis appears to stabilize Co–Co bonding in the chains in a pressure quenchable fashion. **Tables S1** and **S2** summarize our single crystal X-ray diffraction (SCXRD) refinement details.

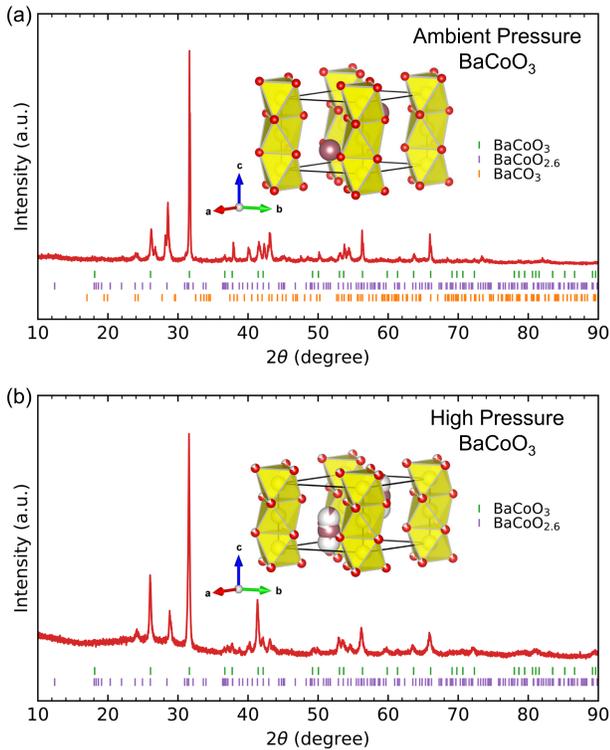

**Figure 1. (a, b)** Powder X-ray diffraction pattern of 2H-BaCoO$_3$ synthesized at ambient pressure and the HP-BaCoO$_3$ material synthesized at 1200 °C and 6 GPa for 3 hrs. Bragg peak positions of each phase are represented by vertical tick marks. The crystal structures are shown in the Insets: Ba, dark red; Co, yellow; O, red.

Diamagnetic behavior is observed down to 2 K in HP-BaCoO$_3$ at low applied fields (10 Oe) as displayed in **Figure 2a**. The magnetic susceptibility measured at 10 Oe starts dropping below zero around 130 K. It can thus be clearly seen that the negative magnetic moments in HP-BaCoO$_3$ can be well described by diamagnetism. The linear Co–Co arrangement along the *c*-axis enhances magnetic interactions between Co ions and magnetic anisotropy, which may be what induces the half-levitation below the transition temperature (~130 K). Thus, the magnetic levitation demonstration was performed in the lab with the ferromagnet cooled by liquid N$_2$, as shown in **Figure 2b**. However, as the applied field is increased to 1000 Oe, ferromagnetism can be seen at a transition temperature (~130 K).

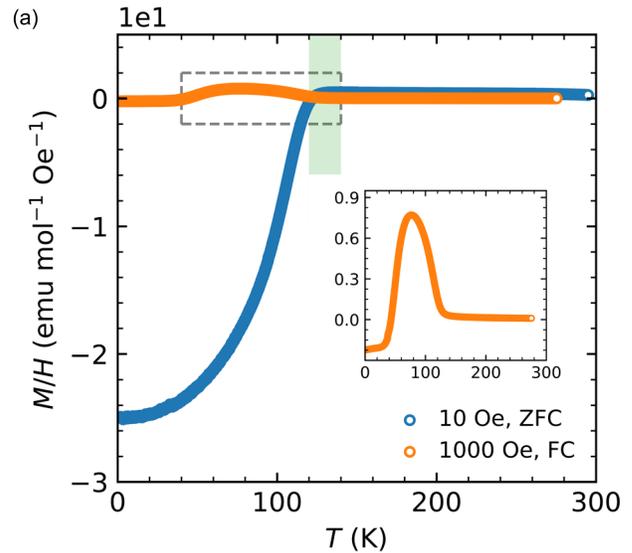

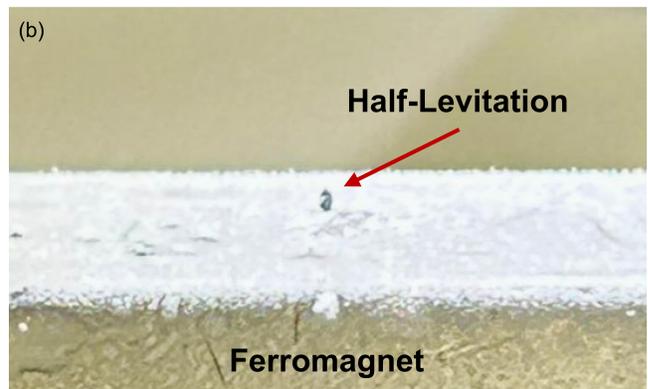

**Figure 2.** Temperature-dependent magnetization study of HP-BaCoO$_3$. **(a)** Magnetic susceptibility measured on heating at 10 Oe and 1000 Oe, after cooling in zero applied field (ZFC). Inset, ZFC data at 1000 Oe. **(b)** Half-Levitation demonstration on HP-BaCoO$_3$ with liquid N$_2$ cooling.

The field-dependent magnetization was measured at 2 K and 100 K and is presented in **Figure 3**. The moment per Co ion, $\mu_{sat}$, was determined to be 0.56(2) $\mu_B$ at 2 K and 9 T, which is smaller than the theoretical value (1.0 $\mu_B$) accounting for a purely localized low spin $d^5$ configuration $S$ = 1/2 Co$^{4+}$. A small Co moment is also observed in other Co-based oxides which may be due to the hybridization between Co and O atoms.[16] It is worth mentioning that the magnetic retentivity in HP-BaCoO$_3$ at 2K is very high, ~0.22(1) $\mu_B$, which indicates the presence of hard ferromagnetic properties for HP-BaCoO$_3$. By 100 K, the magnetic retentivity disappears.

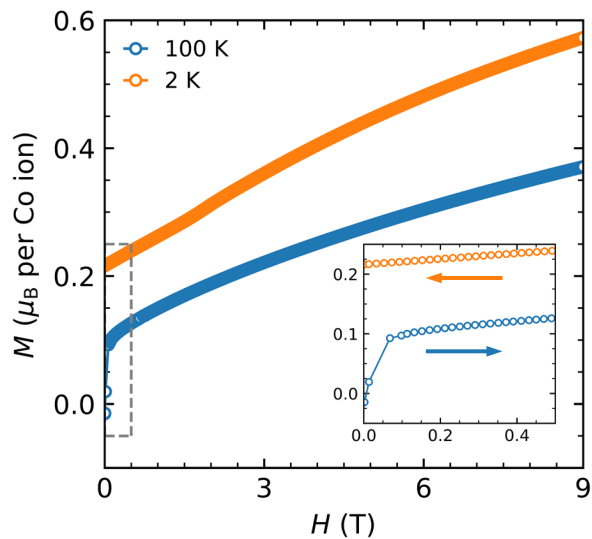

**Figure 3.** The field-dependent magnetization of HP-BaCoO$_3$ at two different temperatures (2 K and 100 K).

The temperature dependence of electrical resistivity of HP-BaCoO$_3$ in the range of 50–300 K is shown in **Figure 4**. The room temperature resistivity is around 0.025(1) Ω cm. It is found that the sample exhibits an insulating behavior at room temperatures and the resistivity increases as the temperature decreases. By around 30 K, the resistivity is too large to be detected by the Physical Property Measurement System (PPMS). The measurement was conducted on two different samples and shows consistent results.

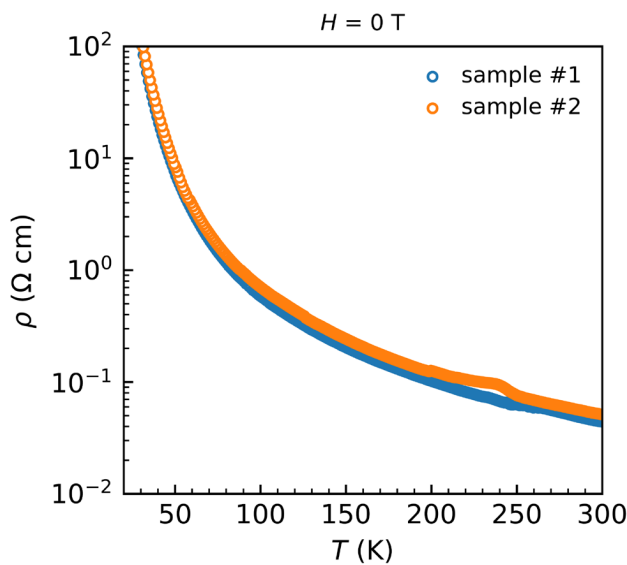

**Figure 4.** The temperature-dependent electrical resistivity of HP-BaCoO$_3$ for two different samples. The small glitch near 250 K seen for one of the datasets is attributed to the presence of water in the measurement system.

We report the high-pressure hexagonal HP-phase of BaCoO$_3$, synthesized at 6 GPa and 1200 °C. Its hexagonal symmetry crystal structure was determined by single crystal X-ray diffraction and is characterized by one dimensional chain of CoO$_6$ octahedra sharing faces with short Co–Co intrachain distances, from which we deduce that Co–Co bonding may be present in the chains. Diamagnetic behavior was observed at low applied magnetic fields, but ferromagnetic behavior is seen at large applied magnetic fields, with a relatively small saturated magnetic moment of 0.56(2) $\mu_B$ that we attribute to $d$–$p$ orbital hybridization. Critically, our electrical resistivity study consistently reveals insulating behavior for HP-BaCoO$_3$. Thus HP-BaCoO$_3$ provides an example that ferromagnetic insulators (we cannot tell at this point whether the material is ferrimagnetic) can impersonate superconductors and even exhibit partial levitation.

## ASSOCIATED CONTENT

**Supporting Information**. Experimental section; TGA analysis; crystal structure and SCXRD refinement of HP-BaCoO$_3$ at room temperature and pressure; atomic coordinates and equivalent isotropic atomic displacement parameters.

## AUTHOR INFORMATION

### Corresponding Author

* Weiwei Xie (xieweiwe@msu.edu)
* Danrui Ni (danruin@princeton.edu)

### Author Contributions

‡ These authors contributed equally.

## ACKNOWLEDGMENT

The work at Michigan State University was supported by the U.S. DOE-BES under Contract DE-SC0023648. The work at Princeton University was supported by the U.S. DOE-BES under Contract DE-FG02-98ER45706.

# Supporting Information

## Impersonating a Superconductor: High-Pressure BaCoO$_3$, an Insulating Ferromagnet

Haozhe Wang[1#], Xianghan Xu[2#], Danrui Ni[2#]*, David Walker[3], Jie Li[4], Robert J. Cava[2], Weiwei Xie[1]*

1. Department of Chemistry, Michigan State University, East Lansing, MI, 48824
2. Department of Chemistry, Princeton University, Princeton, NJ, 08540
3. Lamont Doherty Earth Observatory, Columbia University, Palisades, NY, 10964
4. Department of Earth and Environmental Sciences, University of Michigan, Ann Arbor, MI, 48109

[#] Equally contributed to this work. * Email: xieweiwe@msu.edu; danruin@princeton.edu

## Table of Contents

# Experimental Section

**High-Pressure Synthesis.** The synthesis was conducted using a Walker-type[1] multi-anvil apparatus (MA) at Lamont-Doherty Earth Observatory. The starting material was as-synthesized the ambient pressure $BaCoO_3$ phase, which was prepared by thoroughly mixing the materials $BaCO_3$ and $Co_3O_4$ and subsequently heating them to 900 °C for 12 hrs, then regrinding and reannealing at 1000 °C for 72 hrs.[2] The sample was kept at 120 °C overnight to remove the moisture before high pressure loading. The sample was then loaded in a platinum capsule inside an $Al_2O_3$ crucible that was inserted into a Ceramacast 646 octahedral pressure medium lined on the inside with a $LaCrO_3$ heater and kept at 6 GPa and 1200 °C for 3 hours before quenching to room temperature and then decompressed to ambient pressure overnight (sample serial number TT-1451).

**Phase Analysis.** The phase identity and purity were examined using a Bruker D2 Phaser powder X-ray diffractometer with Cu $K_\alpha$ radiation ($\lambda$ = 1.5406 Å). Room temperature measurements were performed with a step size of 0.010° at a scan speed of 1.33°/min over a Bragg angle ($2\theta$) range of 10–90°.

**Chemical Composition Determination.** The high-pressure synthesis product was examined for purity using a PerkinElmer Thermal Analyzer. The ground powder sample (about 6 mg) was heated in a Pt sample pan under an 5%$H_2$/95%Ar gas flow, and the sequence was: pre-purge for 30 min at room temperature to get rid of extra air; after that, heat up to 1000 °C at 10 °C/min; isothermal at 1000 °C for 2 hrs; and fast cool down to room temperature.

**Structure Determination.** The room temperature crystal structure was determined using a Bruker D8 Quest Eco single crystal X-ray diffractometer, equipped with Mo $K_\alpha$ radiation ($\lambda$ = 0.7107 Å) with an $\omega$ of 2.0° per scan and an exposure time of 10 s per frame. A SHELXTL package with the direct methods and full-matrix least-squares on the $F^2$ model was used to determine the crystal structure of $BaCoO_3$ HP. A SHELXTL package with the direct methods and full-matrix least-squares on the $F^2$ model was used to determine the crystal structure.[3, 4]

**Physical Property Measurement.** Temperature and field-dependent magnetization and electrical resistivity measurements were performed with a Quantum Design DynaCool physical property measurement system (PPMS) under a temperature range of 2–300 K and under applied fields up to 9 T. Electrical resistivity measurements were accomplished with a four-probe method using platinum wires on a polycrystalline sample.

**Figure S1** Thermogravimetric Analysis. The formula obtained was calibrated to Co and listed in the figure.

    a. The weight% vs time(min) curve is shown below (different colors label out different steps in the sequence):

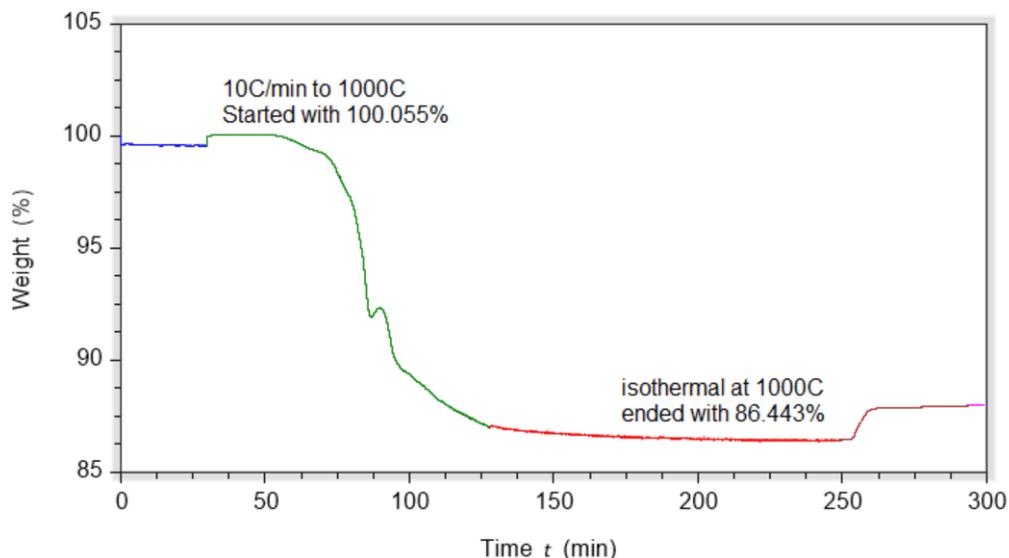

    b. Weight% vs temperature:

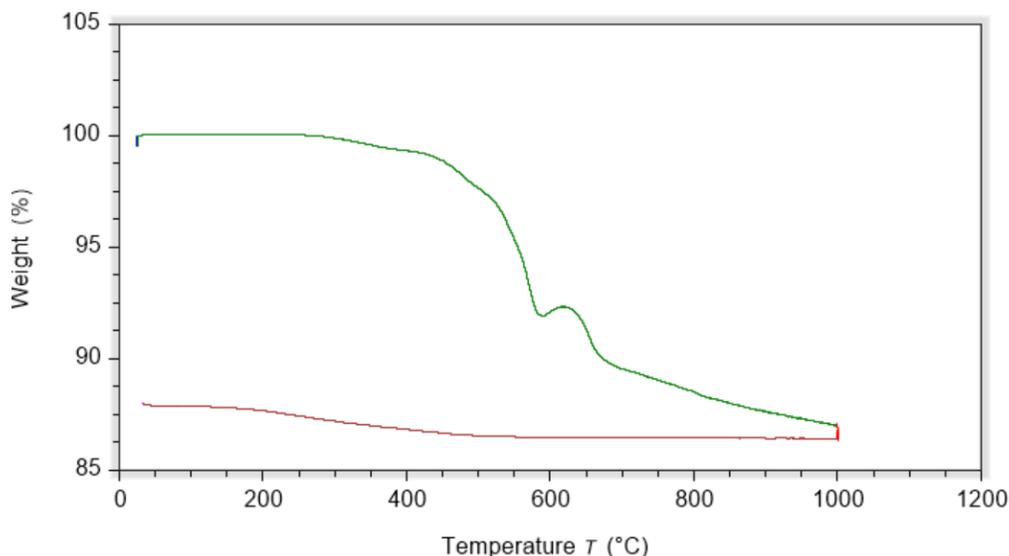

The weight% started with 100.055% and ended at 86.443% after two-hour isothermal heating in 5% $H_2$ in Ar. The reduced product should be BaO + Co, so based on these two numbers the formula will be $BaCoO_{3.092}$. But a small increase was observed when the cooling step started. This can be due to the he sample re-absorbing some moisture or gas from the gas flow or can be some instrumental errors (similar behavior was observed previously on Ba-Co-O-Cl samples reduced in the TGA). If we take the increase into consideration, the weight% when cooling back to r.t. is 87.879%, and the formula is $BaCoO_{2.84}$.

**Table S1** The crystal structure and refinement for HP-BaCoO$_3$ at 300 K.

| Chemical Formula | BaCoO$_{2.37(15)}$ |
|---|---|
| Formula weight | 234.19 g/mol |
| Space Group | $P6_3/mmc$ |
| Unit cell dimensions | $a$ = 5.648(1) Å |
|  | $c$ = 4.138(2) Å |
| Volume | 114.32(7) Å$^3$ |
| Density (calculated) | 7.096 g/cm$^3$ |
| Absorption coefficient | 24.069 mm$^{-1}$ |
| F(000) | 214 |
| $2\theta$ range | 8.34 to 61.38° |
| Total Reflections | 1006 |
| Independent reflections | 80 [$R_{int}$ = 0.0546] |
| Refinement method | Full-matrix least-squares on F$^2$ |
| Data / restraints / parameters | 80 / 0 / 13 |
| Final R indices | $R_1$ (I>2σ(I)) = 0.0452; $wR_2$ (I>2σ(I)) = 0.0835 |
|  | $R_1$ (all) = 0.0541; $wR_2$ (all) = 0.0867 |
| Largest diff. peak and hole | + 0.769 e/Å$^{-3}$ and -0.756 e/Å$^{-3}$ |
| R.M.S. deviation from mean | 0.185 e/Å$^{-3}$ |
| Goodness-of-fit on F$^2$ | 1.343 |

**Table S2** Atomic coordinates and equivalent isotropic atomic displacement parameters (Å$^2$). ($U_{eq}$ is defined as one third of the trace of the orthogonalized $U_{ij}$ tensor.)

| BaCoO$_{2.37(15)}$ | Wyck. | x | y | z | Occ. | $U_{eq}$ |
|---|---|---|---|---|---|---|
| **Co** | 2a | 0 | 0 | 0 | 1 | 0.073(2) |
| **Ba$_1$** | 2d | 1/3 | 2/3 | 3/4 | 0.802 | 0.045(2) |
| **Ba$_2$** | 4f | 1/3 | 2/3 | 0.562(4) | 0.099 | 0.047(6) |
| **O** | 6h | 0.1538(14) | 0.3076(14) | 1/4 | 0.79(5) | 0.148(17) |